# Giant Enhancement of Solid Solubility in Monolayer BNC Alloys by Selective Orbital Coupling

Shiqiao Du[1], Jianfeng Wang[2], Lei Kang[2], Bing Huang[2*], Wenhui Duan[1]

[1] *Department of Physics and State Key Laboratory of Low-Dimensional Quantum Physics, Tsinghua University, Beijing 100084, China and*

[2] *Beijing Computational Science Research Center, Beijing 100193, China*

Email: Bing.Huang@csrc.ac.cn

**Solid solubility (SS) is one of the most important features of alloys, which is usually difficult to be largely tuned in the entire alloy concentrations by external approaches. Some alloys that were supposed to have promising physical properties could turn out to be much less useful because of their poor SS, e.g., the case for monolayer BNC [$(BN)_{1-x}(C_2)_x$] alloys. Until now, an effective approach on significantly enhancing SS of $(BN)_{1-x}(C_2)_x$ in the entire $x$ is still lacking. In this article, a novel mechanism of selective orbital coupling between high energy wrong-bond states and surface states mediated by the specific substrate has been proposed to stabilize the wrong-bonds and in turn significantly enhance the SS of $(BN)_{1-x}(C_2)_x$ alloys. Surprisingly, we demonstrate that five ordered alloys, exhibiting variable direct quasi-particle bandgaps from 1.35 to 3.99 eV, can spontaneously be formed at different $x$ when $(BN)_{1-x}(C_2)_x$ is grown on *hcp*-phase Cr. Interestingly, the optical transitions around the band edges in these ordered alloys, accompanied by largely tunable exciton binding energies of ~1 eV at different $x$, are significantly strong due to their unique band structures. Importantly, the disordered $(BN)_{1-x}(C_2)_x$ alloys, exhibiting fully tunable bandgaps from 0 to ~6 eV in the entire $x$, can be formed on Cr substrate at the miscibility temperature of ~1200 K, which is greatly reduced compared to that of 4500~5600 K in free-standing form or on other substrates. Our discovery not only may resolve the long-standing SS problem of BNC alloys, but also could significantly extend the applications of BNC alloys for various optoelectronic applications.**

# Introduction

Solid solubility (SS) is one of the most critical features for all the alloys, which fundamentally determines their overall physical properties, including electronic [1–3], magnetic [4-5], plasmonic [6-7], defect [8-10] and catalytic properties [11-12], etc. In the past decades, several approaches have been developed to enhance the SS of alloys via the development of various specific growth conditions [6–8,11,13]. In particular, the SS of semiconducting alloys not only depends on their intrinsic structural properties, but also strongly depends on the electronic structures of the host materials [1–5,8–10,14]. The existence of strong phase inhomogeneity, caused by the poor SS, has been a big issue to restrict some semiconducting alloys in their practical applications.

Nitrides have great potentials for solid-state lighting (SSL) [15]. $In_{1-x}Ga_xN$ alloys have appeared to be one of the most important monolithic alloys for white light-emitting-diodes (LEDs). Unfortunately, the poor SS around $x=0.5$ in $In_{1-x}Ga_xN$, induced by the large lattice mismatch (>10%) between InN and GaN, have strongly restricted their efficiency for white SSL [15,16]. Since 2D boron nitride (BN) was discovered as a promising material for deep-UV SSL [17,18], tremendous efforts have been developed to alloying graphene into BN forming BNC solid solutions in order to realize a tunable bandgap ($E_g$) in the entire visible spectrums, so that they might replace $In_{1-x}Ga_xN$ for white SSL [19–31]. However, although the lattice mismatch between BN and graphene is much smaller (<2%) than that of $In_xGa_{1-x}N$, it is widely found that the SS of BNC is even much poorer than that of $In_{1-x}Ga_xN$ [19–31], mostly due to the imbalanced valence electrons of alloyed elements.

In the experiments, ammonia borane ($BNH_6$) molecules are widely adopted as precursors to epitaxially grow BN or BNC alloys [19,32-33], which means that the amount of B and N atoms are mostly equal in the BNC, i.e., forming $(BN)_{1-x}(C_2)_x$, alloys. Although extensive experimental studies have been done on the SS issue of $(BN)_{1-x}(C_2)_x$ [19-27], an effective approach to realize the $(BN)_{1-x}(C_2)_x$ solid solutions in the entire $x$ is still lacking. On the other hand, in the previous theoretical studies [28-31], no clear mechanism was proposed to achieve the ordered or disordered $(BN)_{1-x}(C_2)_x$ solid solutions. Therefore, $(BN)_{1-x}(C_2)_x$ alloys, dominated by the large-scale BN and C domains, mostly behave as very poor electronic properties instead of widely tunable $E_g$ in all the experiments [19–21,23-27].

It is expected that the reduction of structural dimensionality from 3D to 2D could bring new opportunity to modulate the SS of an alloy system. In this article, we have developed an effective idea to overcome the phase inhomogeneity in $(BN)_{1-x}(C_2)_x$. Importantly, a novel mechanism of symmetry-allowed selective orbital coupling between high energy wrong-bond states and specific substrate-mediated surface states has been proposed to stabilize the wrong-bond states and in turn significantly enhances the SS of $(BN)_{1-x}(C_2)_x$ alloys. Surprisingly, we demonstrate that five ordered $(BN)_{1-x}(C_2)_x$ alloys can be spontaneously formed at different $x$ when $(BN)_{1-x}(C_2)_x$ is grown on *hcp*-phase Cr(0001), which have a widely tunable direct quasi-particle $E_g$ from 1.35 to 3.99 eV. Meanwhile, the optical transitions around the band edges in these ordered alloys are significantly strong, which are accompanied by largely tunable exciton binding energies of ~1 eV at different $x$. Moreover, we demonstrate that the calculated miscibility temperature ($T_c$) of $(BN)_{1-x}(C_2)_x$ grown on Cr(0001) can be dramatically reduced from ~5600 to ~1200 K. Once the disordered $(BN)_{1-x}(C_2)_x$ alloys can be formed, they can exhibit a fully tunable $E_g$ from 0 to ~6 eV in the entire $x$.

## Results

**Cluster Expansion Models** The $(BN)_{1-x}(C_2)_x$ system can be effectively considered as a quasi-binary system [29,34], so that the cluster expansion (CE) theory based on first-principles calculations can be applied to calculate the formation energies ($E_f$) of a large number of $(BN)_{1-x}(C_2)_x$ alloys [14, 35]. The basic idea of CE is to expand the energies of a $(BN)_{1-x}(C_2)_x$ configuration into energy contributions of cluster figures (single atoms, pairs, triples, etc.) based on a generalized Ising Hamiltonian:

$$E(\sigma) = J_0 + \sum_i J_i \hat{S}_i(\sigma) + \sum_{i<j} J_{ij} \hat{S}_i(\sigma) \hat{S}_j(\sigma) + \sum_{i<j<k} J_{ijk} \hat{S}_i(\sigma) \hat{S}_j(\sigma) \hat{S}_k(\sigma) + \ldots \quad (1)$$

The index $i$, $j$, and $k$ run over all the alloy sites, and $S_m(\sigma)$ is set to +1 (-1) when it is occupied by BN ($C_2$) dimer. It is noted that the first two terms on the right-band side of Eq. (1) define the linear dependence of the energy of a $(BN)_{1-x}(C_2)_x$ configuration as a function of $x$, while the third and fourth terms contain all pair and three-body interactions, etc. Every cluster figure is associated with an effective cluster interaction (ECI) $J_\alpha$, which indicates the energy contribution of a specific cluster figure to the total energy. Ideally, the CE can represent any $(BN)_{1-x}(C_2)_x$ alloy energy $E(\sigma)$ by the appropriate selection of $J_\alpha$, which can be fitted from first-principles total energy calculations based on a sufficient number of alloy configurations [14]. The cross-validation score is set to 0.05eV and we have further confirmed the convergence of our CE fitting by adding more alloy configurations in the test calculations. To calculate the binary phase diagram of $(BN)_{1-x}(C_2)_x$, Monte Carlo (MC) simulations, which sample a semi-grand-canonical ensemble, are carried out in which the energetics of $(BN)_{1-x}(C_2)_x$ are specified by the CE Hamiltonian [36].

**Phase Inhomogeneity in Free-standing $(BN)_{1-x}(C_2)_x$** Overall, we find that the strong phase separation in $(BN)_{1-x}(C_2)_x$ is mostly due to the high-energy cost of forming B-C and N-C wrong bonds. In BN (graphene), the formation of B-N (C-C) bonds meets the Octet rule that can make the $\pi_{BN}$ ($\pi_{CC}$) bonding states fully-occupied and antibonding $\pi^*_{BN}$ ($\pi^*_{CC}$) states fully-unoccupied, as shown in left (right) panel of Fig. 1. Meanwhile, the large $\pi_{BN}$-$\pi^*_{BN}$ ($\pi_{CC}$-$\pi^*_{CC}$) separation due to the ionic charge transfer (covalent hybridization) during the formation of B-N (C-C) bonds can significantly push the $\pi_{BN}$ ($\pi_{CC}$) states down to a low energy position. As shown in Fig. 1, $\pi_{BN}$ and $\pi_{CC}$ finally reach a similar energy position in terms of our DFT calculations.

During the formation of $(BN)_{1-x}(C_2)_x$, the unavoidable B-C and N-C wrong-bonds do not meet the Octet rule. As shown in Fig. 1 (middle panel), compared to $\pi_{BN}$/ $\pi_{CC}$, $\pi_{BC}$ is pushed down to a higher energy position because the strength of orbital coupling in a C-B bond is weaker than that of B-N/C-C bonds, as indicated by their calculated bond lengths ($d_{C-B}$=1.52 Å, $d_{B-N}$=1.4 Å and $d_{C-C}$=1.42 Å). Meanwhile, $\pi_{BC}$ can only be partially-occupied based on the electron counting. On the other hand, although $\pi_{NC}$ can be pushed down to a lower energy position than that of $\pi_{BN}$/$\pi_{CC}$, due to the stronger N-C orbital coupling strength ($d_{N-C}$=1.38 Å), the high-energy $\pi^*_{NC}$ is unavoidable to be partially-occupied based on the electron counting. In practice, the charge transfer from $\pi^*_{NC}$ to $\pi_{BC}$ may occur to make the $\pi^*_{NC}$ ($\pi_{BC}$) orbitals fully empty (occupied), in order to lower the total energy of the system. Therefore, it turns out that the high energy position of (occupied) $\pi_{BC}$ could be the most critical factor that make the wrong-bonds in $(BN)_{1-x}(C_2)_x$ unstable. Consequently, large BN and C domains always forming in a $(BN)_{1-x}(C_2)_x$, i.e., the less number the wrong-bonds, the lower total energy the $(BN)_{1-x}(C_2)_x$.

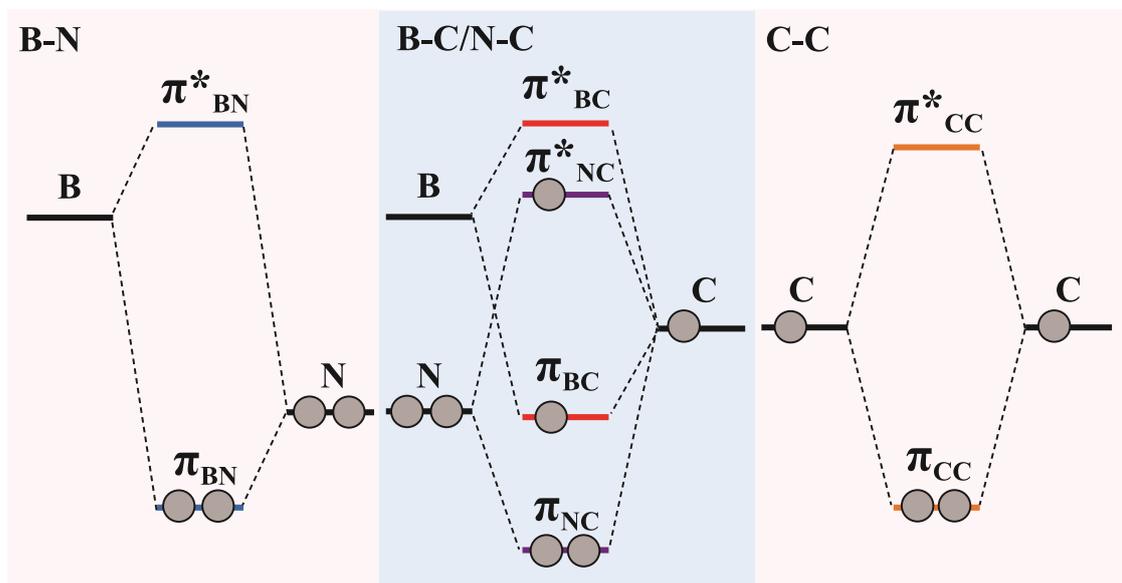

**Figure 1.** Schematic representation of the formations and occupations of bonding $\pi$ and antibonding $\pi^*$ states in B-N (left), B-C/N-C (middle), and C-C (right) bonds, respectively.

The key to enhance SS of $(BN)_{1-x}(C_2)_x$ is to stabilize the B-C and N-C wrong-bonds, i.e., the $\pi_{BC}$ level must be pushed down to a comparable or even lower energy position than that of $\pi_{BN}$/$\pi_{CC}$. In a conventional 3D alloy, it is extremely difficult to modulate the orbital levels of wrong-bonds hidden in the bulk via external approaches. However, it is expected that the reduction of alloy dimensionality from 3D to 2D could bring new opportunity to modulate its SS, as the surface wrong-bond orbitals are touchable and could be further modulated by the external orbital engineering mediated by its surrounding environments. In practice, since $(BN)_{1-x}(C_2)_x$ alloys are usually grown on a variety of transition-metals (TMs), e.g., *fcc*-phase Cu and Ni [19–24, 37], by CVD methods, the external orbital coupling could occur at the surfaces of underneath TMs. Unfortunately, an effective concept on substrate-enhanced SS in $(BN)_{1-x}(C_2)_x$ is still lacking, which is the main purpose of our study.

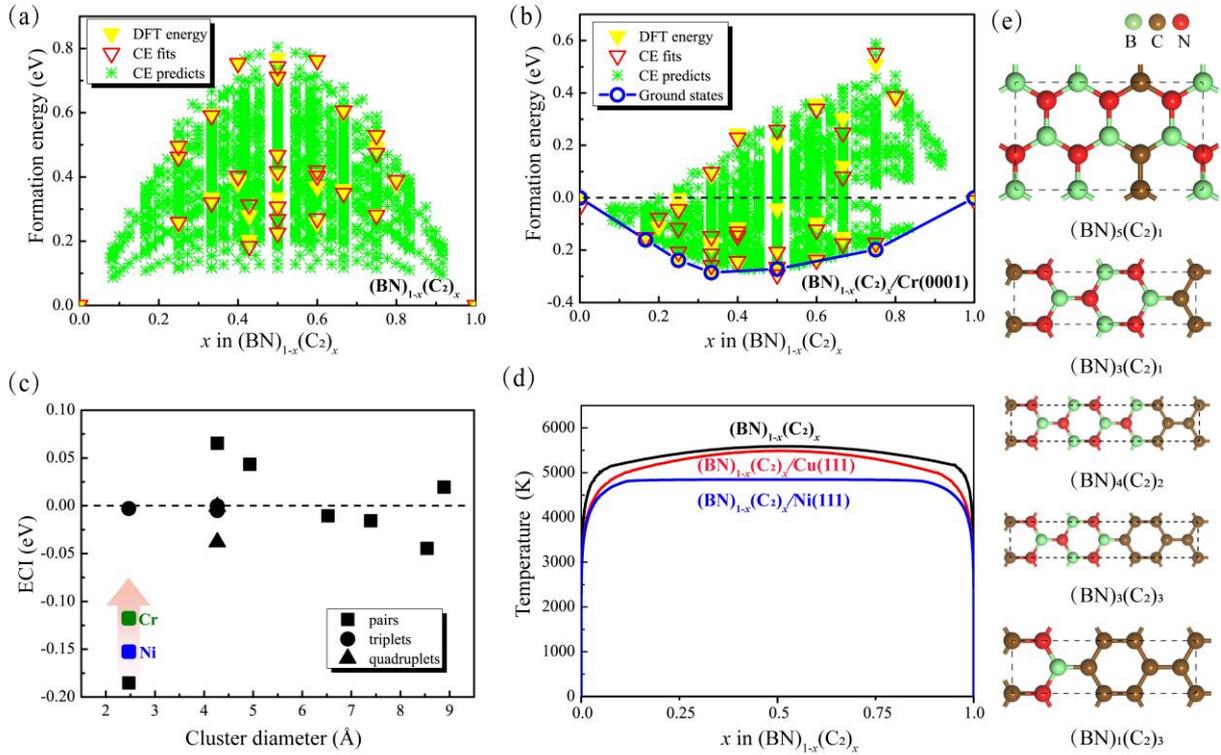

**Figure 2.** Calculated formation energies ($E_f$) of $(BN)_{1-x}(C_2)_x$ (with respect to graphene and BN) along with the corresponding CE fits as a function of $x$ in (a) free-standing $(BN)_{1-x}(C_2)_x$ and (b) $(BN)_{1-x}(C_2)_x$/Cr(0001). The $E_f$ of 2280 and 3020 symmetry-inequivalent $(BN)_{1-x}(C_2)_x$ structures up to 24 atoms/cell calculated from CE are also plotted in (a) and (b), respectively. (c) Effective cluster interactions (ECI) $J_\alpha$ as a function of cluster diameter for free-standing $(BN)_{1-x}(C_2)_x$, fitted with CE. The dominated $J_{\alpha-NN}$ for $(BN)_{1-x}(C_2)_x$/Ni(111) and $(BN)_{1-x}(C_2)_x$/Cr(0001) are also plotted here for comparison. (d) MC-calculated phase diagrams of $(BN)_{1-x}(C_2)_x$ with and without Ni or Cu substrates. (e) Structures of five ordered intermediate $(BN)_{1-x}(C_2)_x$ ground-states in (b), in which the unitcells are enclosed by the dished-lines.

Firstly, we have systematically calculated the $E_f$ of free-standing $(BN)_{1-x}(C_2)_x$, which is defined as

$$E_f = E[(BN)_{1-x}(C_2)_x] - (1-x)\mu_{BN} - x\mu_{C_2} \quad (2)$$

where $\mu_{BN}$ ($\mu_{C2}$) is the energy of BN ($C_2$) unit cell. For free-standing $(BN)_{1-x}(C_2)_x$, the $E_f$ of selected 29 $(BN)_{1-x}(C_2)_x$ structures are calculated using DFT methods, as shown in Fig. 2a. Secondly, these $J_\alpha$ values defining CE are fitting to these 29 DFT energies, and the CE includes 16 $J_\alpha$ up to four-point clusters, as shown in Fig. 2c. Apparently, $J_\alpha$ are dominated by negative values and the largest negative $J_\alpha$ is contributed by the nearest-neighboring ($J_{\alpha-NN}$) BN-BN (or $C_2$-$C_2$) pair attractive interactions, i.e., BN ($C_2$) always prefers to bond with BN ($C_2$) forming large-scale BN (C) domains. Finally, the constructed CE are then used to calculate the $E_f$ of all the enumerated symmetry-inequivalent alloy structures up to 24 atoms/cell. As shown in Fig. 2a. There is no intermediate ground-state for $0<x<1$, i.e., the ground-state of a $(BN)_{1-x}(C_2)_x$ will separate into BN and graphene, and the $E_f$ of metastable $(BN)_{1-x}(C_2)_x$ distribute in a wide range of $0.1<E_f<0.8$ eV, consistent with the dominated negative $J_\alpha$ values [14]. The phase diagram of free-standing $(BN)_{1-x}(C_2)_x$ is calculated by CE-based MC simulations, as shown in Fig. 2d. The miscibility temperature $T_c$ is found to be ~5600K in $(BN)_{1-x}(C_2)_x$, which is too high to be achieved (even higher than the melting point of BN or graphene).

**Phase Inhomogeneity of $(BN)_{1-x}(C_2)_x$ Grown on Cu(111) or Ni(111)** It is curious to further understand the substrate effects on the SS of $(BN)_{1-x}(C_2)_x$. Here the Cu(111) and Ni(111) are selected as representatives not only because they are widely adopted in the current experiments, but also because they are convenient for affordable large-scale alloy calculations with small lattice mismatch (<2%) (See Table 1 in Supporting Materials). For $(BN)_{1-x}(C_2)_x$/TM system, the ground-states of hosts, i.e., BN/TM and C/TM, are first determined by extensive calculations and then the alloy properties of $(BN)_{1-x}(C_2)_x$/TM are calculated following a similar DFT-CE-MC process as that for free-standing $(BN)_{1-x}(C_2)_x$. Overall, the crystal field induced by the $D_{3d}$ symmetry of Cu(111) [or Ni(111)] gives rise to $d$-orbital filling of surface atoms into two double-degenerate states [$e'_g$ ($d_{xy}+d_{x2-y2}$) and $e_g$ ($d_{xz}+d_{yz}$)] and single state $a_{1g}$ ($d_{z2}$). The wavefunctions of $e'_g$, $e_g$, and $a_{1g}$ orbitals are delocalized, and $a_{1g}$ is the only symmetry-allowed $d$ orbital having strong nonzero overlap with the $\pi_{BC}$ level.

For Cu(111), its surface $d$ orbitals are fully occupied (closed shell) and ~5 eV below the Fermi level (See Fig. S1 in Supporting Materials), which cannot couple with $\pi_{BC}$. As a result, the calculated (average) interfacial binding energy ($E_{int}$) between a (metastable) $(BN)_{1-x}(C_2)_x$ and Cu(111) surface is ~62 meV/atom, belonging to weak interactions. No intermediate ground-state is found for 0 <$x$<1 in $(BN)_{1-x}(C_2)_x$/Cu(111) (See Figs. S2-S3 in Supporting Materials). As shown in Fig. 2d, the calculated $T_c$~5300K is slightly lower than that of free-standing $(BN)_{1-x}(C_2)_x$, explaining well the experimental observations of inhomogeneous $(BN)_{1-x}(C_2)_x$ on Cu(111) at a growth temperature of ~1000 K[19–21].

For Ni(111), its surface $d_{z2}$ orbitals are located around the Fermi level, but they are close to be fully (~90%) occupied (See Fig. S1 in Supporting Materials). Therefore, no sufficiently empty $d_{z2}$ orbitals are available to effectively couple with $\pi_{BC}$ lowering its wrong-bond energies. The calculated (average) $E_{int}$ between a (metastable) $(BN)_{1-x}(C_2)_x$ and Ni(111) surface is ~115 meV/atom. Although no intermediate ground-state is found between 0 <$x$<1 (See Figs. S4-S5 in Supporting Materials), the weak $d_{z2}$-$\pi_{BC}$ hybridization in $(BN)_{1-x}(C_2)_x$/Ni(111) can still reduce the strength of $J_{\alpha-NN}$ (Fig. 2c) and in turn slightly reduce its $T_c$ (Fig. 2d). Indeed, the calculated Tc~4850 K for $(BN)_{1-x}(C_2)_x$/Ni(111), which is ~450 K lower than that of $(BN)_{1-x}(C_2)_x$/Cu(111) but still impossible to be achieved[21,37].

**Realization of Ordered $(BN)_{1-x}(C_2)_x$ Alloys on Cr(0001)** Obviously, Cu and Ni, widely adopted in the experiments, are not ideal substrates for growing homogenous $(BN)_{1-x}(C_2)_x$. There are two important criteria for an ideal TM substrate: (1) the surface states hosted by the TM should have sufficient empty $d_{z2}$ orbitals and (2) the energy position of empty $d_{z2}$ orbitals should match well with that of $\pi_{BC}$. In this situation, a key mechanism of selective $d_{z2}$-$\pi_{BC}$ orbital hybridization can occur: as demonstrated in Fig. 3a, the surface empty $d_{z^2}$ orbitals can have a much stronger covalent hybridization with $\pi_{BC}$ than $\pi_{BN}/\pi_{CC}$; after charge transfer from $\pi^*_{NC}$ to $\pi_{BC}$, the fully-occupied $\pi_{BC}$ could be significantly pushed down. This mechanism can effectively stabilize the wrong-bond states in $(BN)_{1-x}(C_2)_x$ and therefore greatly enhance its SS.

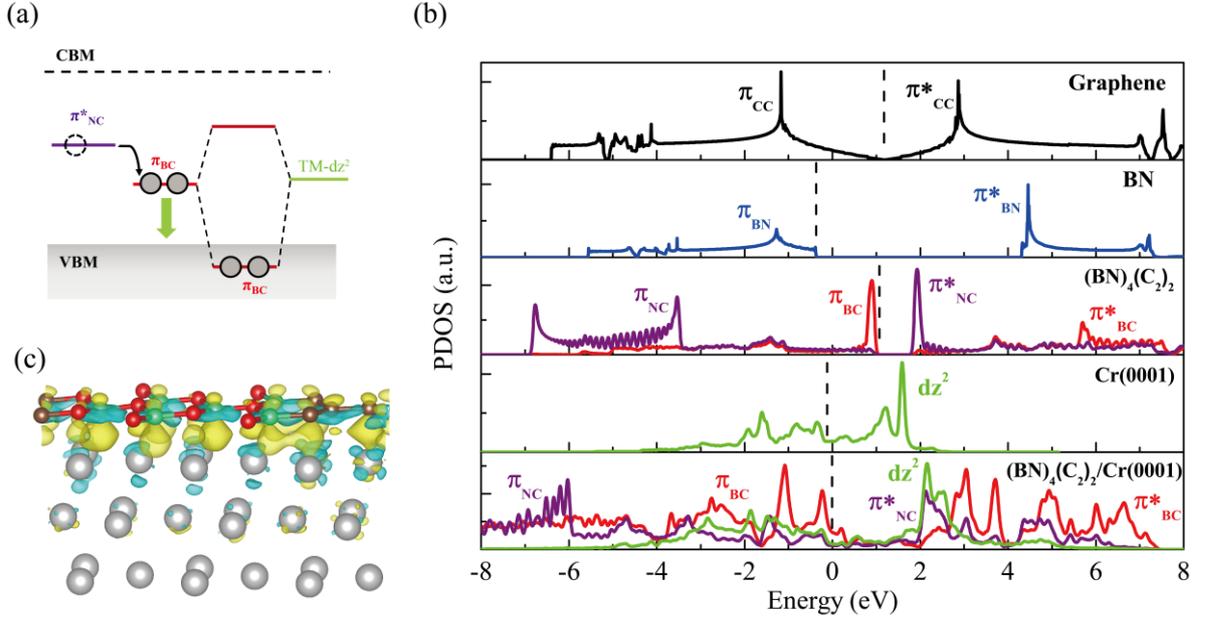

**Figure 3**. (a) Schematic diagram of the mechanism of selective orbital coupling between $\pi_{BC}$ of a $(BN)_{1-x}(C_2)_x$ and $d_{z^2}$ of an ideal TM substrate. (b) DFT-calculated projected density of states (PDOS) of graphene, BN, $(BN)_4(C_2)_2$, Cr(0001), and $(BN)_4(C_2)_2$/Cr(0001). It is noted that only the $d_{z^2}$ orbitals contributed by the Cr surface atoms are plotted here. All PDOS are aligned with the vacuum level and Fermi level positions are marked as dashed lines. (c) Calculated charge density difference for the ground-state $(BN)_4(C_2)_2$ on Cr(0001). The yellow (cyan) areas show where the electron density has been enriched (depleted). The green, brown, red and silver balls represent B, C, N, and Cr, respectively.

Based on the extensive calculations, we successfully discover that the *hcp*-phase Cr, a metastable Cr phase existing in the experiments [38, 39], is such an ideal substrate example for realizing this mechanism. The lattice mismatch between Cr(0001) and $(BN)_{1-x}(C_2)_x$ is also small (<2%) (See

Table 1 in Supporting Materials). As shown in Fig. 2c, the calculated strength of $J_{\alpha-NN}$ is largely reduced in $(BN)_{1-x}(C_2)_x$/Cr(0001) compared to that of free-standing $(BN)_{1-x}(C_2)_x$, which can effectively balance the repulsive and attractive cluster interactions and therefore result in well-ordered ground-states [14]. Interestingly, five intermediate ordered ground-states with specific stoichiometries, i.e., $(BN)_5(C_2)_1$, $(BN)_3(C_2)_1$, $(BN)_4(C_2)_2$, $(BN)_3(C_2)_3$ and $(BN)_1(C_2)_3$, are spontaneously formed in the $(BN)_{1-x}(C_2)_x$/Cr(0001) systems, as shown in Figs. 2b and 2e (See Figs. S6-S7 in Supporting Materials). To further confirm the reliability of our quasi-binary-CE-based structural search, we have considered a more general situation of BNC ternary alloys on Cr(0001) based on the ternary-CE-based structural search approaches [40]. As shown in Fig. S8 in the Supporting Materials, the additional calculations further confirm that the ground-states shown in Fig. 2e are reliable.

For these ground-states on Cr(0001), the calculated average $E_{int}$ is ~210 meV/atom. For Cr(0001), its delocalized surface $d_{z2}$ orbitals, located around Fermi level, are half (~50%) occupied (See Fig. S1 in Supporting Materials), which means that sufficient amount of empty $d_{z2}$ orbitals are available for $d_{z2}$-$\pi_{BC}$ coupling. Taking the ground-state $(BN)_4(C_2)_2$/Cr(0001) as a typical example, the positions of Cr-$d_{z2}$ empty states match well with that of $\pi_{BC}$ in energy, as shown in Fig. 3b. Therefore, the strong $d_{z2}$-$\pi_{BC}$ coupling can significantly broaden, split and push $\pi_{BC}$ down to a much lower energy positions (even lower than that of $\pi_{BN}/\pi_{CC}$ for a large percentage of $\pi_{BC}$ states) and resonant inside the valence band. Meanwhile, it is found that the $d_{z2}$-$\pi^*_{NC}$ orbital coupling can also push the empty $\pi^*_{NC}$ states down below the VBM to be partially occupied and therefore gain energy, as shown in Fig. 3b. However, the orbital coupling strength between $d_{z2}$-$\pi_{BC}$ is significantly stronger than that between $d_{z2}$-$\pi^*_{NC}$. After selective orbital coupling of $d_{z2}$-$\pi_{BC}$ and $d_{z2}$-$\pi^*_{NC}$, the B-C/N-C wrong-bond states can be successfully stabilized in $(BN)_{1-x}(C_2)_x$/Cr(0001).

We have further plotted the charge density difference before and after the $(BN)_4(C_2)_2$ grows on Cr(0001), as shown in Fig. 3c. Interestingly, the largest charge transfer occurs around the region between the B-C wrong-bonds and Cr(0001), which further confirms that the orbital hybridization of $d_{z^2}$-$\pi_{BC}$ is stronger than that of $d_{z^2}$-$\pi^*_{NC}$.

The large bond-length differences between B-C/N-C and B-N/C-C bonds in $(BN)_{1-x}(C_2)_x$ can induce local imbalanced strain fields, which is configuration-dependent and can cost energy. The competition between energy-gain (from orbital coupling) and energy-cost (from the local strains) will effectively determine the structural characteristics [e.g., ultra-narrow BN-C nanoribbon superlattice structures (Fig. 2e)] as well as the amount of wrong-bonds in the $(BN)_{1-x}(C_2)_x$ ground-states. To confirm our conclusion, taking $(BN)_3(C_2)_3$ as a typical example, we have constructed a sufficient number of $(BN)_3(C_2)_3$ configurations that have much more wrong-bonds than the ground-state structure shown in Fig. 2e. Generally, we find that all of these (metastable) structures have larger strain energies and higher $E_f$ than that of the ground-state one (See Fig. S9 and Table 2 in the Supporting Materials). Therefore, the configurations with a larger number of wrong-bonds could have larger structural strain fields and higher $E_f$ compared to that of ground-state configurations.

**Exfoliation of $(BN)_{1-x}(C_2)_x$ from Cr(0001)** The interaction with Cr substrate could make the whole $(BN)_{1-x}(C_2)_x$/Cr system metallic. Therefore, $(BN)_{1-x}(C_2)_x$ monolayer need to be peeled off for practice applications. In experiments, various methods, e.g., wet transfer technique [42,43] or atom/molecule intercalation [44-46], have been developed to exfoliate monolayer materials from different substrates. Here we suggest that the free-standing $(BN)_{1-x}(C_2)_x$ might be obtained by intercalating inert He atoms after the growth process, as shown in Fig. S10 in the Supporting Materials. After growth, the He atoms could be injected into the interfacial space of $(BN)_{1-x}(C_2)_x$/Cr(0001) through the top surface or boundary edges at high temperature with high injection rate. Once He atoms are inserted into the interfacial space, our calculations show that the interfacial bonds between $(BN)_{1-x}(C_2)_x$ and Cr(0001) can be spontaneously broken and then the interfacial distance is significantly enlarged to the vdW range [See Fig. S10 in the Supporting Materials]. In this case, the monolayer $(BN)_{1-x}(C_2)_x$ could be simply peeled off from the Cr(0001) substrate.

Moreover, it is noticed that a new approach, i.e., synthesis-transfer-fabrication process, has recently been developed to realize the device application for monolayer materials with strong substrate interactions, even without the exfoliation process. It has been successfully applied to the silicene/Ag system [47]. It is expected that this approach could also been applied to the $(BN)_{1-x}(C_2)_x$/Cr(0001) system for further device applications.

**Stability of Ordered $(BN)_{1-x}(C_2)_x$ Alloys** We have systemically checked the dynamical and thermal stability of (free-standing) ordered $(BN)_{1-x}(C_2)_x$ monolayers after exfoliation using first-principles phonon spectrum and molecular dynamical (MD) calculations, respectively. As shown in Fig. S11 in the Supporting Materials, no imaginary frequency is found in the phonon spectra of these five ordered $(BN)_{1-x}(C_2)_x$, which means that they are dynamically stable. To confirm the thermal stability of the ground states, the large supercells of >120 atoms are constructed for these five ordered alloys (the same structures used for phonon spectra). The first-principles MD simulations are performed with a Nose-Hoover thermostat at 1000 K. The Fig. S12 in the Supporting Materials shows the fluctuation of total energies of these structures as a function of simulation times. After 6 ps, no structural destruction is found in these structures, except that the large thermal fluctuations can induce rippled structures. Our calculations strongly indicate that the $sp^2$-B-N-C bonds in $(BN)_{1-x}(C_2)_x$ monolayers are stable even under high temperature of (at least) 1000 K.

**Electronic and Optical Properties of Ordered $(BN)_{1-x}(C_2)_x$ Alloys** The quasi-particle band structures of these five ordered $(BN)_{1-x}(C_2)_x$ alloys are calculated based on *GW* calculations (at the $G_0W_0$ level). The excitonic effects of these five ordered $(BN)_{1-x}(C_2)_x$ ground-states are then considered by solving the Bethe-Salpeter equation (BSE) based on their calculated quasi-particle band structures. As a benchmark, the *GW* band structure and optical spectrum of monolayer BN are also calculated (see Fig. S13 in the Supporting Materials), which is in good agreement with the previous calculations [48,49].

As shown in Fig. 4a, all these five ground-states have promising direct gaps ranging from 1.35 to 3.99 eV. The conduction and valence band edges of these band structures are contributed by $\pi^*_{NC}$ and $\pi_{BC}$ states, respectively, which can induce relatively flat band dispersions. It is expected that these flat bands around the band edges can result in high electron-hole pair intensities during either the optical absorption or emission [50].

As a typical example, Fig. 4b shows the calculated absorption spectra of $(BN)_3(C_2)_1$ with electron-hole interaction (*GW-BSE*) and without electron-hole interaction (*GW-RPA*). The optical spectra of other four ordered $(BN)_{1-x}(C_2)_x$ alloys can be found in Fig. S14 in the Supporting Materials. Indeed, the calculated absorption spectra confirm that the optical dipole transitions between $\pi^*_{NC}$ and $\pi_{BC}$ at these band edges are not only allowed but also have much stronger intensities than that of BN. As shown in Fig. 4b, the *GW*-BSE-calculated optical spectrum of $(BN)_3(C_2)_1$ features a strong absorption peak (marked as $A_1$) at ~1.79 eV, which is contributed by the band-edge optical transitions occurring at these *k*-points associated with flat bands, as shown in inset of Fig. 4b.

As shown in Fig. 4c, the *GW* bandgaps ($E_{g\text{-}GW}$) of these ordered $(BN)_{1-x}(C_2)_x$ decrease as *x* increases. Overall, the $E_b$ of these five ordered $(BN)_{1-x}(C_2)_x$ decreases as their $E_{g\text{-}GW}$ decreases, which can been significantly tuned in a very large range of ~1 eV by adjusting *x*. It notes that the exciton binding energy ($E_b$) is determined by the energy difference of the first absorption peak at the absorption edges, e.g., $A_1$ in Fig. 4b, between *GW-PRA* and *GW-BSE* calculations. It is expected that the tunable optical spectra and $E_b$ could have great potential for their optoelectronic device applications.

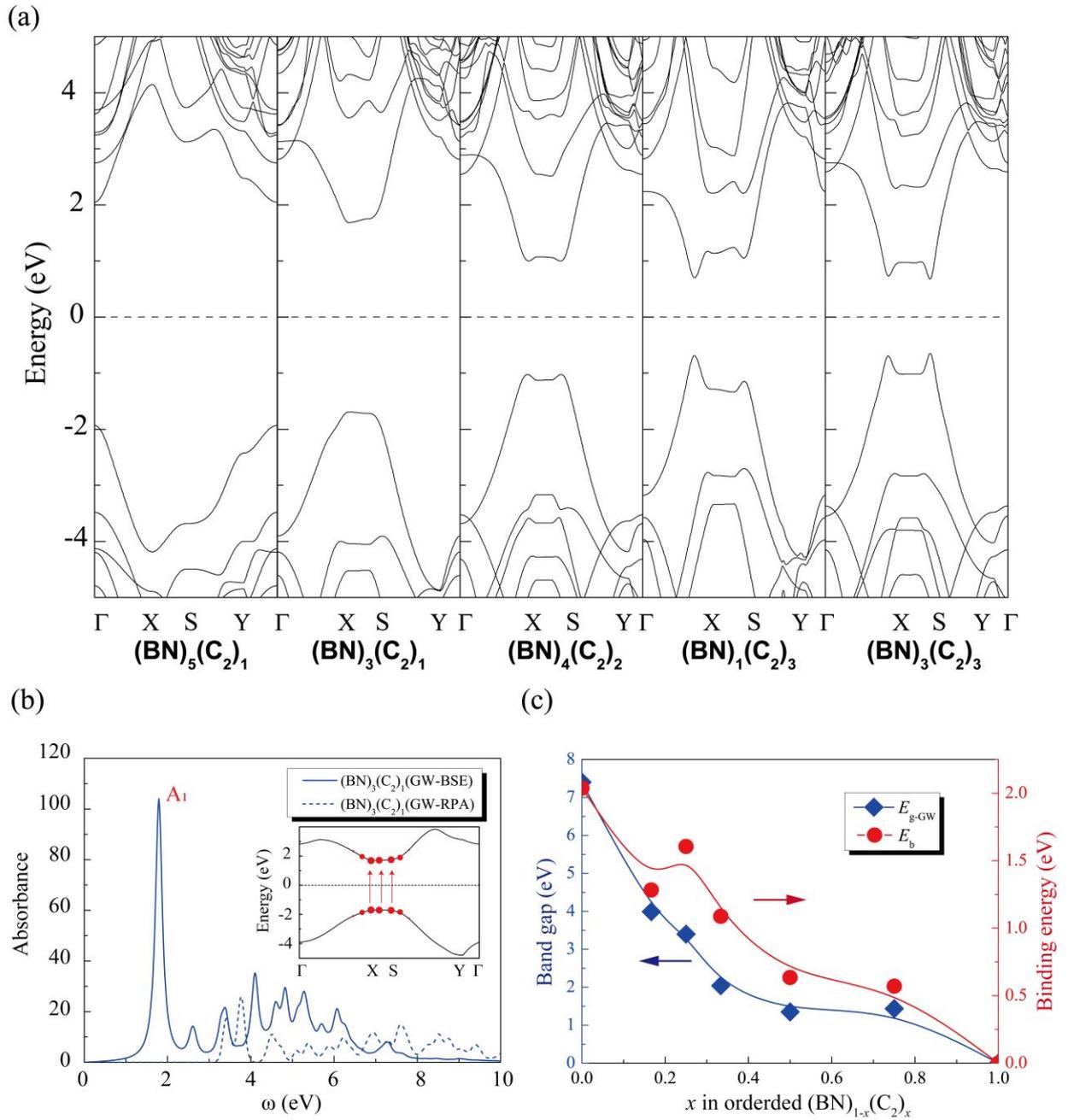

**Figure 4.** (a) *GW* band structures of these five ordered $(BN)_{1-x}(C_2)_x$ with different stoichiometries shown in Fig. 2e. (b) In-plane absorption spectra of $(BN)_3(C_2)_1$ with electron-hole interaction (*GW-BSE*, blue-solid lines) and without electron-hole interaction (*GW-RPA*, blue-dashed lines). Inset shows that the absorption peak (marked as $A_1$) at 1.79 eV is contributed by the band-edge optical transitions occurring at these *k*-points associated with flat bands. (c) Calculated *GW* bandgap ($E_{g\text{-}GW}$) and exciton binding energies ($E_b$) of these five ordered $(BN)_{1-x}(C_2)_x$ together with BN and graphene. The $E_b$ is determined by the energy difference of the first absorption peak at the absorption edges between *GW-PRA* and *GW-BSE* calculations.

**Disordered (BN)$_{1-x}$(C$_2$)$_x$ Alloys** Finally, it is also interesting to investigate the order-disorder phase transition temperatures $T_c$ for these five ordered (BN)$_{1-x}$(C$_2$)$_x$ ground-states on Cr(0001), which can provide a general guideline for the synthesis of disordered (BN)$_{1-x}$(C$_2$)$_x$ at the entire $x$ range. Interestingly, the highest $T_c$ for these ordered configurations is ~1200 K (See Fig. S15-Fig. S19 in the Supporting Materials for the MC simulations), as shown in Fig. 5a. Therefore, it is reasonable to expect that the highest $T_c$ of (BN)$_{1-x}$(C$_2$)$_x$ on Cr(0001) in the entire $x$ is also ~1200 K, as the ground-state configuration at an arbitrary $x$ is always phase separated into a linear combination of its neighboring two ordered ground-state ones. Considering the experimental temperature (>1000 K) on growing (BN)$_{1-x}$(C$_2$)$_x$ [19–24], the homogenous (BN)$_{1-x}$(C$_2$)$_x$ may be successfully achieved in the entire $x$ range, which is a great improvement compared to the previous efforts [19-31]. In order to estimate the $E_g$ of disordered (BN)$_{1-x}$(C$_2$)$_x$ as function of $x$, 6 large-supercell special quasi-random structures (SQS) are selected (See Fig. S20 in the Supporting Materials) [14, 51]. Since it is very challenging to calculate the *GW* band structures of these SQS structures, the HSE06 functional calculations are adopted. As shown in Fig. 5b, the HSE06-calculated $E_g$ of disordered (BN)$_{1-x}$(C$_2$)$_x$ could be continually tuned from 0 to 5.7 eV (See Fig. S21 in Supporting Materials). Our calculations strongly indicate that (BN)$_{1-x}$(C$_2$)$_x$ could be the first monolithic alloy systems that can cover the entire energy range from infrared to deep-UV spectrums, as long as they can be grown on Cr or other similar functional substrates in the future experiments.

## Conclusion

In conclusion, a novel mechanism of selective orbital coupling between wrong-bond states and surface states mediated by the substrates has been proposed to significantly enhance the SS of 2D (BN)$_{1-x}$(C$_2$)$_x$ alloys. Based on this mechanism, we have discovered that ordered and disordered solid solutions could be obtained when (BN)$_{1-x}$(C$_2$)$_x$ is grown on Cr(0001). Our discovery not only resolves the long-standing SS problem of BNC alloys, but also could significantly extend their great potentials for electronic and optoelectronic applications.

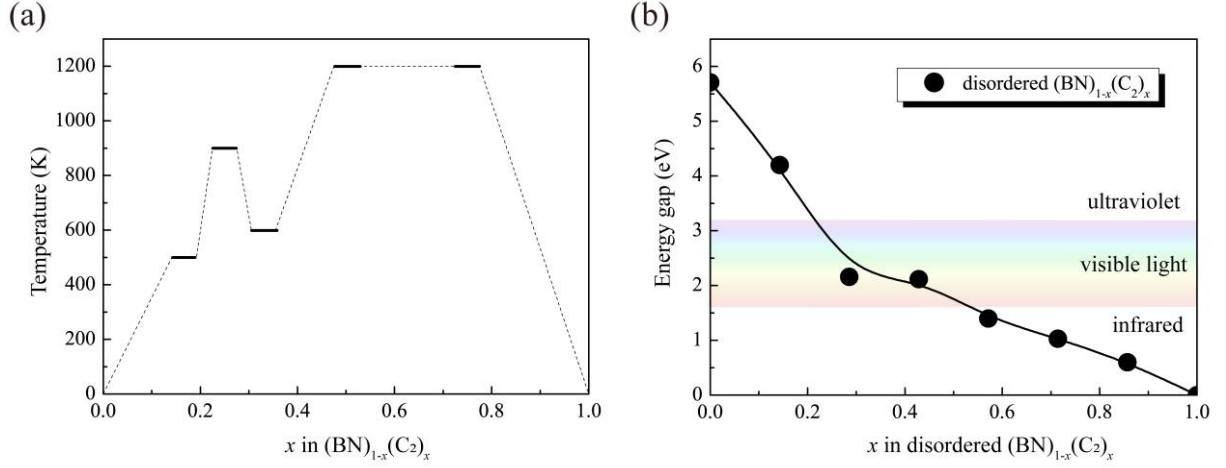

**Figure 5**. (a) MC-simulated order-disorder phase transition temperatures for these five intermediate ordered $(BN)_{1-x}(C_2)_x$ alloys grown on Cr(0001). (b) HSE06-calculated bandgaps of disordered $(BN)_{1-x}(C_2)_x$ alloys as a function of $x$.

## Methods

All the first-principles density functional theory (DFT) calculations are preformed using VASP package with the GGA-PBE for exchange correlation functional [52]. The cut-off energy for plane wave basis is set to be 520eV. A $\Gamma$-centered $k$-point grid that is generated automatically (2000 $k$-points per reciprocal atom) to make the mesh as uniform in the reciprocal space for different alloy structure. The convergence criterions of electronic step iteration for structural relaxation and a static run after the relaxation are $10^{-5}$eV and $10^{-6}$eV respectively. The vacuum layer is chosen for > 12 Å to eliminate the spurious periodic errors along the $z$ direction. A fixed volume relaxation method is applied for freestanding $(BN)_{1-x}(C_2)_x$ structures. For $(BN)_{1-x}(C_2)_x$/TM (TM=Cu, Ni, Cr) structures, the $(BN)_{1-x}(C_2)_x$ alloys are fully optimized on the fixed cleaved three layers TM substrates. The DFT-D3 method is applied to include the van de Waals (vdW) interactions [53]. All the structures are relaxed until the force on each atom is less than 0.01eV/ Å. To accurately estimate the band gaps, the hybrid functional (HSE06) [54] is applied to accurately estimate the electronic structures of disordered $(BN)_{1-x}(C2)_x$ alloys.

For CE simulation, we have modified ATAT [35] code in order to effectively calculate the quasi-binary $(BN)_{1-x}(C2)_x$ alloy systems. For the MC simulations, a semi-grand-canonical ensemble is sampled on a superlattice that can contain a sphere with 35 Å radius until the averaging and equilibration time are reached and then phase boundary tracing technique is used to determine the boundary efficiently [36].

For the electronic and optical calculations of ordered $(BN)_{1-x}(C2)_x$ alloys, the eigenvalues and wavefunctions that are obtained by standard first-principles DFT calculations are adopted for *GW* calculations at one-shot $G_0W_0$ level. The number of bands ($N_b$) and energy cutoff (ECUTGW) in *GW* calculations are tested carefully to ensure bandgaps well-converged to <0.1eV. The Wannier90 code package is used to generate band structures based on the *GW* calculations. The Bloch functions are projected onto *s* and *p* orbitals of B, C and N respectively. The energy windows are set about from -5eV below to +5eV above the Fermi level in which the first-principles calculations are exactly reproduced by the Wannier90 fitting. The Bethe-Salphter equation (BSE) is solved finally to obtain optical absorption spectra with the consideration of excitonic effects in the basis of previous *GW* calculations.

For the special quasi-random structures (SQS) simulations, six $(BN)_{1-x}(C2)_x$ SQSs (98 atoms per supercell) are generated by MC simulations to calculate the electronic structures of fully disordered $(BN)_{1-x}(C2)_x$ alloys at different *x*.

## Acknowledgements

We thank Dr. S. -H. Wei at CSRC for the helpful discussions. S.D. and J. W. contribute equally to this work. We acknowledge the support from NSFC (Grant Nos. 11574024, 11674188 and 51788104), MOST of China (Grant No. 2016Y-FA0301001), NSAF U1530401 and the Beijing Advanced Innovation Center for Materials Genome Engineering. The calculations were performed at Tianhe2-JK at CSRC.